\documentclass[paper,twocolumn,twoside]{geophysics}
\usepackage{multicol}
\usepackage{float}
\usepackage{graphicx}
\usepackage{amssymb}
\usepackage{csquotes}
\usepackage{lipsum}
\usepackage{csquotes}
\usepackage{color}

\usepackage{amsbsy}
\usepackage{amsmath}
\usepackage{cleveref}
\usepackage{soul}

\creflabelformat{equation}{#2#1#3}

\begin{document}
	
	\let\oldref\ref
	\renewcommand{\ref}[1]{(\oldref{#1})}
	\title{Transdimensional 2D Full-Waveform Inversion and Uncertainty Estimation}
	
	\renewcommand{\thefootnote}{\fnsymbol{footnote}}

	\address{
		\footnotemark[1]Institute for Geophysics, \\
		John A. and Katherine G. Jackson School of Geosciences \\
		The University of Texas at Austin \\
		Austin, TX 78713-8924\\
		E-mail: reetam@utexas.edu; mrinal@utexas.edu \\
	}
	\author{Reetam Biswas\footnotemark[1]  and Mrinal K. Sen\footnotemark[1]}

	\righthead{Trans-dimensional 2D FWI using RJHMC}
	
	
	\newpage
	
	\begin{abstract}
			
		Full-Waveform Inversion (FWI) has now become a widely accepted tool to obtain high-resolution velocity models from seismic data. It restores the high-frequency features in a velocity model using both the time and amplitude information from the seismic data. Typically, the velocity model in its discrete form is represented on a rectangular grid, which is uniformly distributed throughout the model space, and we solve for the elastic properties at these grid points. FWI is mostly solved employing a local optimization method, where one obtains a velocity update by minimizing the misfit between the observed and the calculated seismograms. Note also that FWI is a highly non-linear problem which is known to be prone to non-uniqueness. The convergence to a globally optimum solution is not guaranteed; it depends on the choice of the starting model. Thus, a Bayesian formulation of the inverse problem with subsequent sampling of the posterior distribution is a preferred choice, since it enables uncertainty quantification. However, with the increase in the dimension of a model, sampling search space becomes computationally expensive. We employ a recently developed trans-dimensional sampling method called Reversible Jump Hamiltonian Monte Carlo (RJHMC), to the 2D full waveform inversion problem. We represent our velocity model using Voronoi cells, determined from the distribution of certain nuclei points in the model space. This method offers two advantages. First, it solves for a variable dimensional velocity updates by using a trans-dimensional reversible jump Markov Chain Monte Carlo (RJMCMC) step and thus tries to achieve an optimum number of nuclei to represent the model and minimize the misfit. A smaller number of parameters helps in an efficient sampling of the model search space. Second, it applies the gradient-based Hamiltonian Monte Carlo (HMC) step, which further improves the sampling by allowing the algorithm to take a large step guided by the gradient. This two-step algorithm proves to be a useful tool for model exploration and uncertainty quantification in FWI. 
	\end{abstract}

	\section*{Introduction}
	\renewcommand{\figdir}{fig}

	Unlike travel-time tomography and Amplitude versus offset (AVO),  FWI \citep{tarantola1986strategy,virieux2009overview} uses both amplitude, and phase information to generate subsurface velocity models using an iterative least-square data-fitting procedure. It minimizes the data misfit by updating the velocity model with a scaled gradient of the misfit function. Generally, the adjoint state method~\citep{lailly1983seismic,tarantola1984inversion,pratt1990inverse} is used to calculate the gradient of the misfit function. One of the most straightforward formulations of FWI is the constant-density acoustic FWI, which solves for the P-wave velocity only. 
	
	Commonly, we use a pre-determined number of model unknowns in a geophysical inversion problem, namely, the number of layers or cells to represent a velocity model. There exist several statistical techniques to understand the choice of model parametrization required for optimal data misfit like F-tests~\citep{aster2011parameter}. The above formulation requires the FWI problem to have a fixed dimension of the model parameters. However, the subsurface properties vary greatly with space, and thus the sensitivity of seismic observation also varies spatially. In FWI, an over-parameterized model leads to poor conditioning of the inverse problem, and is mostly tackled by introducing regularization. The regularization term damps the solution and smooths the model~\citep{sen2003computation,virieux2009overview}. In geophysical problems, fixing the number of parameters may lead to two cases. Starting an inversion with too few parameters will underfit the data, while on the other hand, starting with too many parameters will overfit the data \citep{sen1991nonlinear}. Both of these conditions lead to erroneous parameter estimation with extreme uncertainty~\citep{dosso2014efficient}. A trans-dimensional formulation of the problem helps in overcoming these shortcomings.

	The problem of determining the model dimension during inversion, i.e., treating the number of parameters as a parameter itself, has recently received some attention in the geophysical community. Solving such an inverse problem may seem unrealistic. Note that a Bayesian sampling framework has a built-in parsimonious character; a high-dimensional model is naturally discouraged \citep[e.g.,][]{malinverno2002parsimonious}. Statisticians have actively developed several trans-dimensional Markov chain  Monte Carlo (MCMC) methods that allow transition between states having different dimensions, i.e., simultaneously traversing both parameter and model space \citep[for a  review, see][]{sisson2005transdimensional}. Out of these, the most commonly used is the reversible jump algorithm (RJMCMC) of \cite{GREEN01121995} \citep[see also][]{green2001delayed,sambridge2006trans} for exploring variable dimensional models. \citet{betancourt2017conceptual} showed that the parameter sub-space becomes more and more complex as the parametrization moves towards the higher dimensional space, i.e., increasing the number of parameters, the tension between the density and volume increases exponentially and the region from where significant contribution could come becomes narrower. Using a trans-dimensional approach helps to eradicate some of these drawbacks. It shows a parsimonious character and tries to represent the model using the optimal number of required parameters, keeping the model simple (enough such that it is neither under nor over-parameterized); thus, achieving better exploration and uncertainty quantification of model space in comparison to the same model with more parameters and using a fixed-dimensional inversion scheme \citep{dettmer2016tsunami,olugboji2017statistical}.
	
	\cite{malinverno2000seg} first applied the reversible jump algorithm to the problem of inversion of a zero-offset vertical seismic profile~\citep[for  a complete treatment see][]{malinverno2005monte}. Subsequently, \cite{malinverno2002parsimonious} applied it to an electrical-resistivity inversion. Later, \cite{agostinetti2010receiver} extended the work to solve  for velocity models using receiver-function data. \cite{bodin2009seismic} applied it to a seismic tomography problem using Voronoi cells. \cite{ray2012bayesian} apply trans-dimensional inversion to marine  controlled-source electromagnetic (CSEM) data, and later \cite{ray2016frequency} solved a frequency-domain 1D elastic FWI problem. \cite{dosso2014efficient} applied  RJMCMC to Geo-acoustic profile estimation; \cite{dadi2016velocity}  for velocity log up-scaling using RJMCMC. Even though RJMCMC has better model exploration than traditional fixed-dimensional Bayesian inversion, it faces the problem of having low model acceptance ratio. \citet{al2004improving} proposed a hybrid technique to combine RJMCMC with some other method to improve the model acceptance probability. Most recently, \citet{reetamtrans} combined RJMCMC with the Hamiltonian Monte-Carlo (RJHMC) to speed up convergence and applied it to 1D seismic inversion. \citet{hawkinstrans} used HMC as one of the cases alongside the birth-death scheme; each gets triggered by some probability one at a time.
	
	Using HMC can boost the convergence as well as the acceptance ratio of RJMCMC; \citet{reetamtrans} show a comparison of using a conventional RJMCMC with the RJHMC  for a 1D seismic inversion problem. Although RJHMC converges in fewer iterations, it comes at the cost of additional forward problems to calculate the gradient since it uses gradient information to generate a model update and minimize the problem of sampling correlated models which most Bayesian techniques face. Recently, HMC has gained popularity in solving geophysical problems \citep[for e.g., see][]{reetamtrans,fichtner2018hamiltonian,hawkinstrans}. However, to achieve a good sampling in HMC, it needs to be tuned, which is somewhat problem-dependent. \citet{betancourt2017conceptual}, and \citet{fichtner2018hamiltonian} discuss different possible ways to tune the Mass matrix in HMC to have a minimum correlation between the samples and explore the model space.
	
	One other limitation of FWI is the requirement of a good starting model. Although this problem has recently been addressed using global optimization methods \citep[e.g.,][]{datta2016estimating}, no attempt has yet been made to characterize uncertainty using a fully nonlinear sampling method. \citet{hejunZhu} estimate uncertainty assuming the posterior probability distribution (PPD) to be Gaussian and starting with a model obtained from a local optimization as the prior. \citet{kotsi20184d} used Bayesian framework to address uncertainty in estimating velocity changes in 4D FWI. In this paper, we primarily address these two issues using a trans-dimensional approach. We treat the model parametrization and velocity values as unknown in the inversion  and estimate the uncertainty. Here, we apply RJHMC to a computationally intensive 2D FWI problem and demonstrate its feasibility and usefulness using a noisy synthetic dataset.

	\section*{Theory}
	
	To solve the FWI problem, we first describe the parametrization used in describing a given velocity model and then the  forward modeling algorithm. Finally, we describe the inverse formulation for model exploration based on RJHMC.
	
	\subsection{Model Parametrization} 
	
	To solve a trans-dimensional 2D FWI problem using RJHMC, we choose to represent our model space using an adaptive ensemble of nuclei. The 2D P-wave velocity model is partitioned using a  variable number of non-overlapping Voronoi tessellation~\citep{okabe1992spatial,bodin2009self}. The Voronoi nuclei constitute parametrization of the model space, and each Voronoi cell encloses all the points in space which are closer to the nucleus of that Voronoi cell than any other Voronoi nuclei. Here, we treat the number and position of these Voronoi nuclei as unknown, thus making the Voronoi cell mesh evolving along with the nucleus. The Voronoi nuclei determine the elastic property distribution in the model. The nuclei spatial distribution, as well as their number, can adaptively change as a function of iteration during sampling to an efficient number of parameters so as to match the synthetic data with the observed data and reduce the misfit. These may lead to concentrating nuclei where high-resolution is required in the model, and sparser mesh  elsewhere~\citep{agostinetti2015local,bodin2012,sambridge1995geophysical,agostinetti2015local}.

	To demonstrate an example, \Cref{fig:voronoiexample} shows a seismic velocity model discretized using a set of Voronoi polygons. Let us consider a set $\textbf{C}$ of $k$ nuclei in the 2D plane ($\textbf{C} = {c_1,..., c_k}$ where $c_i\in \mathbb{R}^2$ ). The Voronoi tessellation defines $k$ mutually exclusive regions $R_1$, $R_2$,...,$R_k$ such that the points in the area which fall within the region $R_i$ are closer to $c_i$ than other $c_j (j \ne i)$. Note that the cell nuclei may not necessarily represent the geometric center. The boundaries between the cells  represent the perpendicular bisectors of each pair of neighboring nuclei. Thus in a 2D space, each nucleus $c_i$ is defined by its position ($x$ and $z$ coordinates) and a  seismic wave velocity $v_i$. Therefore, the model parameters having $k$ number of nuclei can be represented by 
	
	\begin{equation}
	\mathbf{m}=(k,\mathbf{C},\mathbf{V}),
	\label{eqn:rjhmcfwi1}
	\end{equation}
	where $\mathbf{V}$ is the vector of the velocity values assigned to each nuclei. Thus the effective number of unknowns (dimension of the model) is $3k$ (for each nucleus, it's X and Z position, and its velocity). During the inversion, the number of nuclei, their position, and elastic property are variable and thus the generated Voronoi cell can take different size, shape, and velocity. This partition modeling approach directly leads to a dynamic parametrization and makes the data determine the cell geometry and the grid resolution to achieve a maximum resolution power of data.

	To generate the full-scale 2D fine mesh grid velocity model from the nuclei, we first create the Voronoi map between the nuclei; the grid points which fall within the Voronoi area are assigned a velocity from that respective nucleus. In \Cref{fig:voronoiexample}, the white cross-hairs (does not represent all the grid points in the figure for simplicity) show the fine mesh grid laid on top of the Voronoi map. Using Voronoi cells as a parametrization may have a discontinuous border for neighboring cells having different velocities, therefore a smoothing is performed to minimize artifacts due to the edge reflection.
	
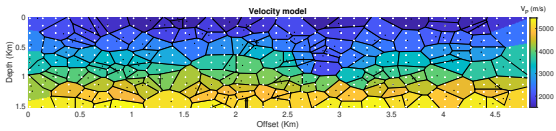
	\plot{voronoiexample}{width=\columnwidth}{Figure showing the velocity model created from nuclei cells represented in black dot. The white cross does not represent all the grid points in the figure for simplicity are the grid points in the velocity model. The color in the Voronoi cells are the velocity of that respective nuclei in that area.}

	\subsection{Forward Modeling} 
	
	Once a gridded model is obtained, we perform forward modeling using the  constant-density acoustic wave equation, where the only elastic parameter is the P-wave velocity. Such type of wave equation is given by
	\begin{equation}
	\frac{1}{c^2}\frac{\partial^2P}{\partial t^2}=\nabla^2P+S(x,t),
	\label{eqn:rjhmcfwi2}
	\end{equation}
	where $P$ is the pressure wavefield, $c$ is the P-wave velocity, $\nabla^2$ is the Laplacian given by $\frac{\partial^2}{\partial x^2}+\frac{\partial^2}{\partial z^2}$, and $S(x,t)$ is the source term. 
	
	At each iteration, after the velocity model is perturbed, the wave equation is solved to compute new seismogram ($\mathbf{d}_{cal}$) and the misfit from the observed seismogram ($\mathbf{d}_{obs}$) is calculated using
	\begin{equation}
	E=\|\mathbf{d}_{obs}-\mathbf{d}_{cal}\|_2.
	\label{eqn:rjhmcfwi3}
	\end{equation}

	The gradient is calculated by the adjoint-state method~\citep{lailly1983seismic,tarantola1984inversion,pratt1990inverse}, where the data residual is backward propagated in reverse time from the receivers to calculate the adjoint wavefield by
	\begin{equation}
	\frac{1}{c^2}\frac{\partial^2R}{\partial t^2}=\nabla^2R+\Delta S(x,t),
	\label{eqn:rjhmcfwi4}
	\end{equation} 
	where $R$ is the adjoint wavefield, and $\Delta S(x,t)$ is the data residual between the observed data ($\mathbf{d}_{obs}$) and the calculated data ($\mathbf{d}_{cal}$). Then the gradient of the misfit function is calculated by summing the zero-lag cross-correlation of the two wavefields over multiple shots 
	\begin{equation}
	\frac{\partial E}{\partial m}=\frac{1}{c(x,z)^3}\sum_{shots}\ddot{P}(x,z,t)R(x,z,t),
	\label{eqn:rjhmcfwi5}
	\end{equation}
	where P and R are the source and adjoint wavefields in space and time.
	
	\subsection{Trans-dimensional Bayesian Formulation}
	
	We employ a Bayesian approach to full waveform inversion with the goal of estimating multiple solutions for uncertainty quantification. Choice of prior probability distribution is crucial to sample the regions near the global minimum efficiently \citep{tarantola1984inversion,buland2003bayesian,bosch2010seismic,Karimi2010}. In a Bayesian framework~\citep{mosegaard1995monte}, information about the model parameters is expressed by a probability  distribution that describes the associated uncertainties, known as the posterior probability distribution (PPD). It represents the intersection of the set of possible solutions, which satisfy our prior knowledge, and our observations. The PPD can be represented by the Bayes rule as follows:
	
	\begin{equation}
	\hat{\pi}(\mathbf{m})=\frac{p(\mathbf{m})p(\mathbf{d|m})}{p(\mathbf{d})},
	\label{eqn:rjhmcfwi6}
	\end{equation}
	
	where $p()$ is a probability distribution, $A|B$ indicates conditional dependence between $A$ and $B$, and $\textbf{d}$ denotes the calculated data. Thus, $\hat{\pi}(\textbf{m})$ is the PPD, $p(\textbf{m})$ represents the prior information available  on the model $\textbf{m}\in\mathbb{R}^{k}$ ($k$ is generally assumed to be known), and $p(\textbf{d}|\textbf{m})$ is the likelihood (also represented as $l(\textbf{m})$), which is the probability as to how well the synthetic data generated by the given model $\textbf{m}$ matches the observed data $\textbf{d}$.  The denominator $p(\textbf{d})$ is called the \enquote{evidence}  or the marginal likelihood~\citep{tarantola1982generalized,hirata1987maximum,cary1988automatic,sen1996bayesian,bernardo2001bayesian}. It is defined as the numerator on the right-hand side integrated over the entire model space and thus regarded as a constant. Therefore, we can write \Cref{eqn:rjhmcfwi6} in the form of a proportionality relationship:
	
	\begin{equation}
	\hat{\pi}(\mathbf{m})\propto p(\mathbf{m})p(\mathbf{d|m}).
	\label{eqn:rjhmcfwi7}
	\end{equation}
	
	We use Markov Chain Monte Carlo (MCMC) type of method to sample models from the PPD using \Cref{eqn:rjhmcfwi7}. The new model ($\textbf{m}'$) is sampled from an assumed proposal distribution ($q(\textbf{m}'|\textbf{m})$) and is accepted using Metropolis-Hastings acceptance criterion as
	
	\begin{equation}
	\alpha(\textbf{m}'|\textbf{m})=\min\Bigg[1,\frac{q(\textbf{m}|\textbf{m}')}{q(\textbf{m}'|\textbf{m})}\frac{p(\textbf{m}')}{p(\textbf{m})}\frac{\ell(\textbf{m}')}{\ell(\textbf{m})}\Bigg].
	\label{eqn:rjhmcfwi8}
	\end{equation}
	
	The above MCMC algorithm draws samples from a fixed number of model parameters. The most commonly used trans-dimensional MCMC method is the reversible jump MCMC algorithm. In RJMCMC, the random walk can jump between the states having different number of unknowns. Solving a trans-dimensional problem needs some modification to the Bayes' rule \citep{malinverno2000seg} to accommodate the variability in the model parametrization as
	
	\begin{equation}
	\pi(k,\textbf{m})=\frac{p(k)p(\textbf{m}|k)p(\textbf{d}|\textbf{m},k)}{\sum_{k\in\kappa}\int p(k)p(\textbf{m}|k)p(\textbf{d}|\textbf{m},k)dm},
	\label{eqn:rjhmcfwi9}
	\end{equation}
	where $\textbf{m}\in \mathbb{R}^{c\times k}$ represents a model in the $M_k$ dimensional parameter search space associated with $k$ layers. In our FWI problem, $c=3$. Here, $p(k)p(\textbf{m}|k)$ represents the combined prior probability of the state $(k,\textbf{m})$ (assuming independent  prior), which is the product of prior for the model dimension $p(k)$ and the prior for the model parameter values $p(\textbf{m}|k)$.  The function $p(\textbf{d}|\textbf{m},k)$ represents the likelihood of the state $(k,\textbf{m})$,  also written as $\ell (\textbf{m})$. The integration in the denominator represents  the normalization factor (also known as evidence) in case of a trans-dimensional problem. It  integrates out the product of prior and the likelihood for all possible values of $k\in\kappa$, where $\kappa$ represents the different number of parametrization cells possible. Similar to \Cref{eqn:rjhmcfwi7}, we can represent it as
	
	\begin{equation}
	\pi(k,\mathbf{m})\propto p(k)p(\mathbf{m}|k)p(\mathbf{d|m},k).
	\label{eqn:rjhmcfwi10}
	\end{equation}
	
	\cite{GREEN01121995} defines the acceptance ratio for accepting a new sampled model $(\textbf{m}',k')$ given the current model $(\textbf{m},k)$ as
	
	\begin{equation}
	\alpha(k',\textbf{m}'|k,\textbf{m})=\Bigg[1,\frac{\pi(\textbf{m}',k')q(\textbf{m},k)}{\pi(\textbf{m},k)q(\textbf{m}',k')} |\textbf{J}|\Bigg].
	\label{eqn:rjhmcfwi11}
	\end{equation}
	This equation shows the transition from model $\textbf{m}\in\mathbb{R}^{c\times k}$ (dimension of $c\times k$) to a model $\textbf{m}'\in\mathbb{R}^{c\times k'}$ (dimension of $c\times k'$). $q(\textbf{m})$ is the proposal distribution used in sampling a new model. Here, $|\textbf{J}|$ represents the determinant of the Jacobian matrix, which takes into account a change in the dimension. The Jacobian term acts as a normalization factor for the change in volume between two different dimensional spaces.
	
	\section{Reversible Jump Hamiltonian Monte Carlo (RJHMC)}
	
	Here, we follow  the Reversible Jump Hamiltonian Monte Carlo algorithm~\citep{reetamtrans}, which is a stochastic method based on Bayes' rule. It consists of two steps~\citep{al2004improving}. The first step is a trans-dimensional step similar to RJMCMC. In this step, the model $\mathbf{m}$ with $k$ nuclei may jump towards an increase in dimension to the proposed state $\mathbf{m}'$ having $k'=k+n$ nuclei. The prior for the model dimension is given by $p(\mathbf{m}',k')$. The second step is a fixed-dimensional Hamiltonian Monte Carlo (HMC)~\citep{neal2011mcmc}, where the model $\mathbf{m}'$ moves to a new state $\mathbf{m}^*$ having the same number of nuclei points $k'$. Note that in the reverse step of this move, we first perform the fixed dimensional HMC to change the state of the model from $\mathbf{m}^*$ to $\mathbf{m}'$ having $k'$ number of nuclei and then perform the trans-dimensional step to move model $\mathbf{m}'$ having $k'$ nuclei to model state $\mathbf{m}$ with $k$ nuclei. The RJHMC acceptance probability gives the transition from model $\mathbf{m}$ $(k)$ to model $\mathbf{m'}$ $(k')$ to final model $\mathbf{m}^*$ $(k')$.

	\subsubsection{First: Trans-dimensional step (RJMCMC)}
	
	For a first trans-dimensional step, the random moves can jump between the states having different model dimensions (number of nuclei here). Here, we have the transition from model $\textbf{m}$ with $k$  nuclei to model $\textbf{m}'$ having $k'$ nuclei. We have adopted the birth-death type of scheme \citep{agostinetti2015local} to carry out the trans-dimensional step. This scheme consists primarily of three different moves. In each of these moves, a candidate model is generated by randomly perturbing the current model following the given proposal distribution. The three moves are as follows:
	
	\begin{enumerate}
		\item \emph{No Perturbation}: In this move, we do not perturb the model space and keep it the same. This step has a half probability.
		\item \emph{Birth}: We add $n$ number of nuclei at a randomly chosen available locations using a prior probability distribution. The new nucleus is added by keeping in mind that the distance between any two nearby nuclei is greater than the minimum distance allowed. This step has a quarter probability.
		\item \emph{Death}: We remove $n$ number of nuclei randomly from the current nuclei. This step also has a quarter probability.
		
	\end{enumerate}
	
	The acceptance ratio for a Bayesian inversion dealing with change in dimension has a Jacobian matrix as described in \Cref{eqn:rjhmcfwi11}. Now, for a birth-death type of approach described above, the determinant of this Jacobian matrix can be shown to be unity (please refer to Appendix~B in \citet{reetamtrans} for full detail), i.e., $|\textbf{J}|=1$. Now we will describe the various distributions used in the trans-dimensional part of the algorithm.
	
	\paragraph{Prior Distribution ($p(\textbf{m})$):}
	
	Prior probability in a Bayesian algorithm represents our prior knowledge about the model. The distribution can be in the form of some complex formulation to a simple bound applied to the model. Here, we assume that the two parameters (number of nuclei, and velocity) in the model $\textbf{m}$ are not correlated \textit{a priori} so that the complete prior distribution can be represented as a product of the prior distribution of the two:
	
	\begin{equation}
	p(\textbf{m})=p(k)p(\textbf{m}|k).
	\label{eqn:rjhmcfwi12}
	\end{equation}
	
	The prior on the number of parameters (partitions) $p(k)$ is chosen to be a bounded uniform distribution between the minimum number of nuclei $k_{min}$ and the maximum number of nuclei possible $k_{max}$, represented as:
	
	\begin{gather}
	p(k)=
	\begin{cases}
	\frac{1}{({k_{max}-k_{min}})} \quad   \text{if } k_{min}\le k\le k_{max}\\
	0 	\quad \quad \quad \quad\quad\quad\quad\quad	\text{Otherwise}.\\
	\end{cases}
	\label{eqn:rjhmcfwi13}
	\end{gather}

	The prior distribution on the parameter values ($p(\mathbf{m}|k)$) is defined as a Gaussian distribution as
	\begin{eqnarray}
	p(\mathbf{m}|k)=\frac{1}{\big[(2\pi)^k \det(\mathbf{\bar{C}}_p)\big]^{1/2}}\exp\Bigg[-\frac{1}{2}(\mathbf{m}-\mathbf{\bar{m}})^T(\mathbf{\bar{C}}_p)^{-1}\nonumber\\(\mathbf{m}-\mathbf{\bar{m}})\Bigg],
	\label{eqn:rjhmcfwi14.5}
	\end{eqnarray}
	where $\mathbf{\bar{C}}_p$ is the prior covariance matrix and $\mathbf{\bar{m}}$ represents the prior mean model.

	\paragraph{Likelihood Function ($p(\textbf{d}|\textbf{m},k)$):}
	
	This plays a major role in the acceptance probability of \Cref{eqn:rjhmcfwi9}. This term quantifies the measure of error between the predicted data and the observed data. In general, we assume the error between the observed and the predicted data to be normally distributed and is quantified by a covariance matrix $\textbf{C}_{err}$. It takes into account various factors like noise in the data, approximation error in modeling, etc. Estimating $\textbf{C}_{err}$ requires extensive work and is not considered in our study. Here, we assume a simple case in which the error between different data points are uncorrelated and thus can represent the covariance matrix as a diagonal matrix of the same standard deviation $\sigma_{err}$, i.e., $\textbf{C}_{err}=\sigma^2_{err}\textbf{I}$, where $\textbf{I}$ is an identity  matrix. Thus, the likelihood function $p(\mathbf{d}|\textbf{m},k)$ in \Cref{eqn:rjhmcfwi9} can be represented as 
	\begin{equation}
	p(\textbf{d}|\textbf{m},k)=\frac{1}{\big[(2\pi\big)\sigma_{err}]^{N/2}}\exp\Big[-\frac{1}{2}\Big(\frac{E}{\sigma_{err}}\Big)^2\Big].
	\label{eqn:rjhmcfwi15}
	\end{equation}

	\subsubsection{Second: Fixed-dimensional step (HMC)}

	In this section we describe a fixed Dimensional MCMC step, where the model in state $\textbf{m}'$ with $k'$ nuclei moves to a new state $\textbf{m}^*$ with same $k'$ nuclei using Hamiltonian Monte Carlo. \cite{duane1987hybrid} proposed this method based on an approximate simulation of Hamiltonian dynamical system using the potential energy given by log-target density. It helps in avoiding issues with high curvature and concentration of measures. It  produces a distant proposal for the Metropolis algorithm and avoids slow exploration by using gradient information while changing states \citep{betancourt2017conceptual} . In HMC, we calculate the total potential energy $(U)$ and kinetic energy $(K)$ of the state, i.e., the Hamiltonian Energy. HMC augments the state space of the target distribution by adding auxiliary sets of variables termed as momentum $\textbf{p}$. Each nucleus is augmented with an initial momentum value derived from a normal distribution $N(0,1)$. The total Hamiltonian energy of the model $\textbf{m}$ with momentum $\textbf{p}$ can be given by
	\begin{equation}
	H(\textbf{m}^*,\textbf{p})=U(\textbf{m}^*)+K(\textbf{p}).
	\label{eqn:rjhmcfwi16}
	\end{equation}
	
	The potential energy $U$ is defined as the negative logarithm of the posterior distribution given as 
	\begin{equation}
	U(\textbf{m}^*)=-\log\big[p(\textbf{m}^*)p(\textbf{d}|\textbf{m}^*)\big],
	\label{eqn:rjhmcfwi17}
	\end{equation}
	where the prior probability $p(\textbf{m}^*)$ is taken to be a uniform bounded distribution and the likelihood $p(\textbf{d}|\textbf{m})$ is same as the \Cref{eqn:rjhmcfwi15}. The kinetic energy term $K$ is calculated by the equation given as
	\begin{equation}
	K(\textbf{p})=\frac{\textbf{p}^T\textbf{M}^{-1}\textbf{p}}{2},
	\label{eqn:rjhmcfwi18}
	\end{equation}
	where $\textbf{M}$ is the mass matrix. \citet{fichtner2018hamiltonian} have emphasized the critical role of mass matrix in the tuning of the HMC sampling for a fixed-dimensional problem. It can be used to put different weights to different parameters and control their perturbations and achieve sampling with most uncorrelated samples. In our trans-dimensional case, since the number of parameters is continuously changing with iterations, we chose an identity matrix. To perform the Hamiltonian step, the model $(\textbf{m}',k')$ at the beginning of the second step is assigned a momentum $(p)$ value for each nucleus. It has a Hamiltonian energy $H(\textbf{m}',\textbf{p})$. This momentum value is drawn randomly from a normal distribution $N(0,1)$. Then, we simulate Hamiltonian dynamics $L$ times. We have used the Leapfrog technique to simulate the Hamiltonian dynamics~\citep{VanGunsteren1988,neal2011mcmc}. In each Hamiltonian step, we update the momentum $p$ with the help of gradient information for a $\delta/2$ time step. The gradient is calculated using the FWI adjoint-state method. We use this half updated momentum to update the model for the complete $\delta$ time step, and again using the gradient calculated from this new model, we update the momentum for a $\delta/2$ time step, such that after this one loop both momentum and model have traversed a complete $\delta$ time step. At the end of the Hamiltonian simulation, we have a new state of model $\textbf{m}^*$ and momentum $\textbf{p}^*$ with Hamiltonian energy $H(\textbf{m}^*,\textbf{p}^*)$. We calculate the difference in the total Hamiltonian energy of the two states. Note that the parameters $L$ and $\delta$ have to be carefully tuned for specific applications. The updates for the position state and the momentum vector in each iteration of HMC thus can be given by
	\begin{eqnarray}
	p_i^{new}&=&p_i^{old}-\frac{\delta}{2}\times\frac{\partial U}{\partial m_i},\nonumber\\
	\textbf{m}_{new}&=&\textbf{m}_{old}-\delta\times\textbf{p}^{new},\nonumber\\
	p_i^{new}&=&p_i^{old}-\frac{\delta}{2}\times\frac{\partial U}{\partial m_i^{new}}.
	\label{eqn:rjhmcfwi19}
	\end{eqnarray}

	\subsubsection{Acceptance Ratio for RJHMC}

	After completion of the two steps, i.e., a transition from a model  $(\textbf{m},k)\rightarrow(\textbf{m}',k')\rightarrow(\textbf{m}^*,k')$, we accept this new model with an RJHMC acceptance probability given in ~\citet{reetamtrans}. Note that in the case of a birth step, we first have a trans-dimensional step followed by an HMC step. However, in the reverse move, we first go through the HMC step followed by a trans-dimensional death step. This is to maintain reversibility of the chain. Now, for the above transition, we can write the detail balance equation as 
	\begin{eqnarray}
	\int_{(\textbf{m},\textbf{m}',\textbf{m}^{*})\in A\times B\times C}\nonumber\\
	\pi(\textbf{m})q(\textbf{m}',{k'})P(\textbf{m}^{*}|\textbf{m}')\alpha(\textbf{m}^{*},k'|\textbf{m},k)d\textbf{m}d\textbf{m}'d\textbf{m}^{*}\nonumber\nonumber\\
	=\int_{(\textbf{m},\textbf{m}',\textbf{m}^{*})\in A\times B\times C}\nonumber\\
	\pi(\textbf{m}^{*})P(\textbf{m}'|\textbf{m}^{*})q(\textbf{m},{k})\alpha(\textbf{m},k|\textbf{m}^{*},k')d\textbf{m}d\textbf{m}'d\textbf{m}^{*},
	\label{eqn:rjhmcfwi20}
	\end{eqnarray}
	for all Borel sets $A \in \mathbb{R}^{c\times k}$, $B \in \mathbb{R}^{c\times k'}$, and $C \in \mathbb{R}^{c\times k'}$. \citet{reetamtrans} give the acceptance probability for the above inverse formulation. The acceptance probability is
	\begin{eqnarray}
	\alpha(\textbf{m}^{*},k'|\textbf{m},k)=&\min\Bigg[1,\frac{p(\textbf{m}^{*}|k')p(\textbf{d}|\textbf{m}^{*},k')}{p(\textbf{m}|k)p(\textbf{d}|\textbf{m},k)}\times\nonumber \\ &\frac{q(\textbf{m}|k)}{q(\textbf{m}'|k')}\times\frac{\exp(-H(\textbf{m}'))}{\exp(-H({\textbf{m}^{*}}))}\Bigg].
	\label{eqn:rjhmcfwi21}
	\end{eqnarray}
	where $q(\textbf{m}'|k')$ is the proposal distribution for the model parameter values that is carefully chosen to be an approximation of a posterior distribution, to achieve a high value of acceptance ratio. \Cref{fig:flowchart} shows the flowchart of the algorithm.
	
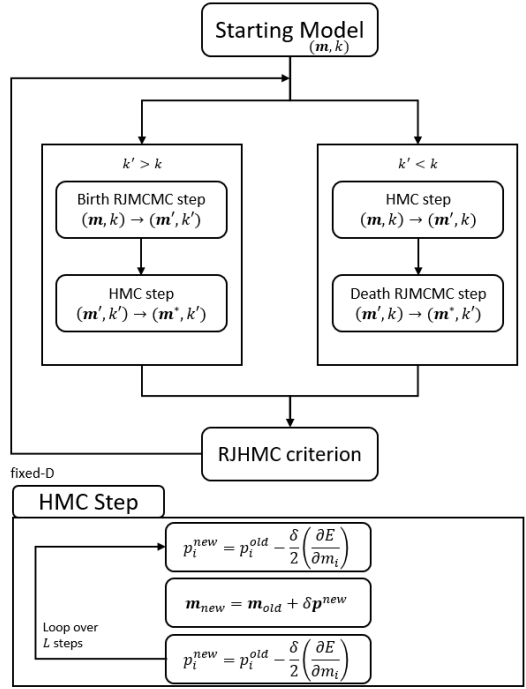
	\plot{flowchart}{width=0.8\columnwidth}{Flowchart of RJHMC.}

	\section{Result}
	
	We demonstrate our algorithm using a synthetic dataset generated using the Marmousi model for P-wave velocity. The model has dimensions of $251$ (grid points in depth) $\times 767$ (grid points in the horizontal direction), with a grid spacing of 12 m. A 15 Hz Ricker wavelet was used to generate the synthetic data set. We used 21 shots at an interval  of $36$ grid spaces with 767 receivers per shot. Gaussian noise was added to the traces, which were then used in the inversion. \Cref{fig:seismogram} shows a shot gather for one of the shots. We started with 1000 nuclei randomly distributed in our model space, and we performed two steps of the leap-frog Hamiltonian Dynamics for each iteration of RJHMC. In the trans-dimensional step of RJHMC, the algorithm can add or delete up to a maximum of 30 nuclei at a time.
	
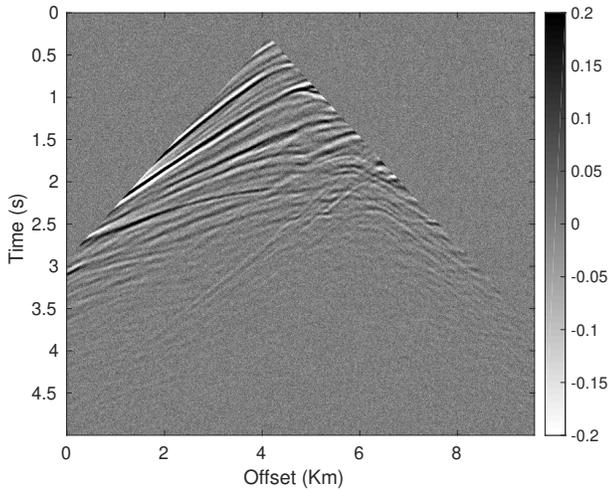
	\plot{seismogram}{width=\columnwidth}{Figure shows one of the generated shot gathers. Note that the direct arrivals have been muted.}
	
	Before applying the RJHMC algorithm to the Marmousi model, we performed a test on the prior distribution. A dry run was made with no input data while setting the likelihood function to be unity to sample the model space. With the absence of a likelihood function, the algorithm should produce a posterior identical to the prior distribution. For this test, we set $k_{max}$ to be 20000. \Cref{fig:kPrior} represents almost a uniform distribution, i.e., all possible values of nuclei count within the  bound of $k$ are possible, similar to the defined uniform distribution in \Cref{eqn:rjhmcfwi13}. Similarly, \Cref{fig:modelPrior} represents prior mean values of the P-wave velocity as defined in the prior for the model parameter as in \Cref{eqn:rjhmcfwi13}. These have a close resemblance with the prior distribution defined and thus we can conclude that the algorithm is sampling correctly. 
	
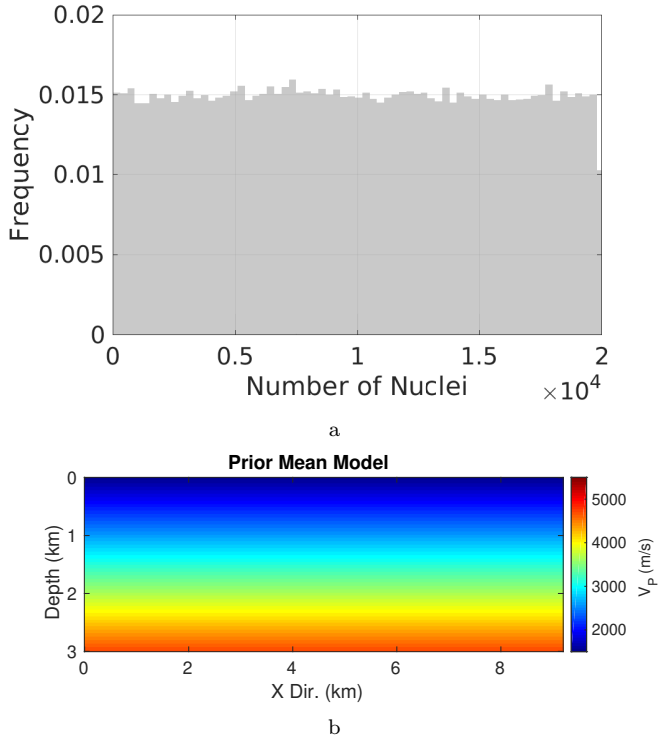
	\multiplot{2}{kPrior,modelPrior}{width=\columnwidth}{(a) Figure showing the distribution of number of nuclei, and  (b) Figure showing the mean velocity model with likelihood function set to unity.}

	To throw light on the convergence of the algorithm, we plot the normalized error curve and the running mean plot of the P-wave velocity at five randomly chosen locations. The error is generally high at the initial iterations, since, the starting model can be quite far away from most likely models. However, with the successive iteration of MCMC, current state reaches closer to the target distribution \citep{betancourt2017conceptual} and samples from there; at this point, the error stabilizes. This initial part of exploration is called the burn-in phase and the models after this phase samples the posterior distribution and thus is used for calculating the uncertainty in model parameters. We ran a total of 15000 iterations of RJHMC, considering the computation complexity of the problem.  \Cref{fig:error15k} shows the normalized error progression along with iteration. By examining the error curve, we note that there is a stability in error at around 5000 iterations. We  also generated a running mean of $V_P$ at five randomly chosen locations in \Cref{fig:runningMean}. All five plots show excellent convergence. Based on these two plots, we infer the completion of burn-in phase to be around 5000 iterations.

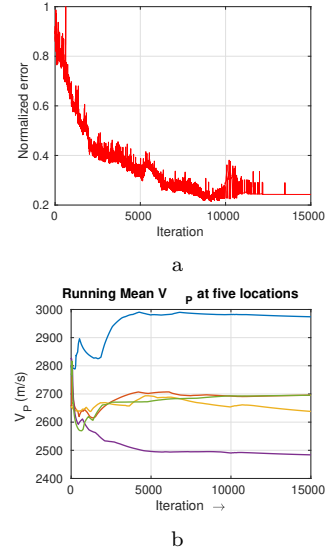
	\multiplot{2}{error15k,runningMean}{width=0.5\columnwidth}{(a) Plot of error curve along with iteration (b) Plot of the running mean $V_P$ parameters for 5 randomly chosen locations.}

	The MCMC method performs poorly in a high-dimensional problem. With the increase in dimension, the sample space narrows down significantly and thus becomes more difficult to sample \citep{betancourt2017conceptual}, and the consecutive samples become more and more correlated. This makes the posterior biased and limited to the vicinity of the starting model. To test the capability of RJHMC, we calculate the autocorrelation of the successive models sampled during the run. \Cref{fig:corr} represents the correlation matrix for 2500 successive models starting from the $5000^{th}$ iteration (after burn-in phase). The matrix shows a high value near the diagonal and minimal values close to zero at the far end. This shows that sample independence in RJHMC has been achieved with successive iterations. \Cref{fig:autocorrelation} shows the auto-correlation coefficient of the starting model with the successive models up to iteration 2500. This figure shows  how rapidly the correlation coefficient of the model falls to a small value.

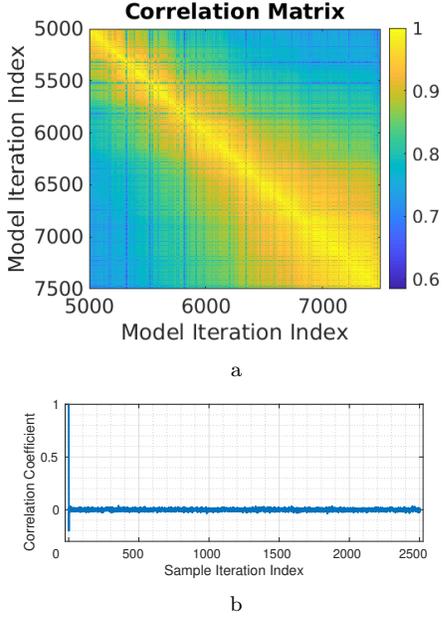
	\multiplot{2}{corr,autocorrelation}{width=0.7\columnwidth}{(a) Plot of the correlation matrix using the models from iteration 5000 to 7500 (2501 models)  (b) Plot of the correlation coefficient of starting model with the successive models at various iterations up to iteration 2500.}

	\Cref{fig:modelswithIter} shows the progress of the model during sampling. The first model is the true velocity model.  We used a smooth starting model  drawn from the prior and sampled with 1000 nuclei  as shown in \Cref{fig:modelswithIter}b. The models in the other panels show the velocities at various iterations of RJHMC along with the nuclei count at that step. We started with 1000 nuclei as a staring model (\Cref{fig:modelswithIter}b). Note that, in general, we observe an increase in the number of nuclei until convergence. \Cref{fig:modelswithIter}d shows the model from the last iteration.
	
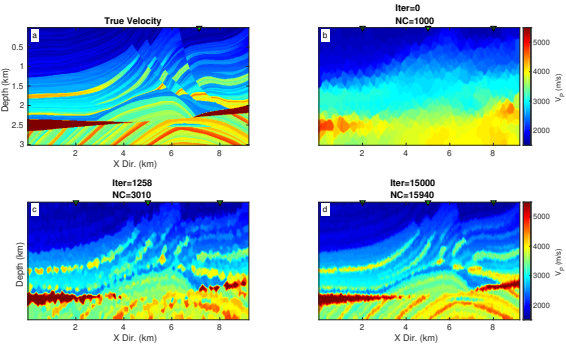
	\plot{modelswithIter}{width=\columnwidth}{Figure showing velocity models at different iterations of RJHMC: (a) the true velocity model, (b)-(d) models at progressive iterations. Plots show the iteration number and the number of nuclei count in the model represented as NC.}

	\Cref{fig:locationPPD2_1} shows the velocity models at $X=(2,5,8)$ km.  It shows the plot of the marginal PPD (grey cloud), true log (green), mean parameter estimate (red), and 95\% confidence interval (dashed line) for the wells marked with green location. The well has the target reservoir of the Marmousi model at a depth of 2.3 km that is fairly well resolved.  Comparison of the Posterior distribution at the three well locations in \Cref{fig:locationPPD2_1} shows that the true model is almost everywhere within one posterior standard deviation of the posterior mean model. However, in the deeper part of the model, the posterior uncertainty is closer to the prior.  \Cref{fig:kPPD15k1} shows the marginal posterior density for the number of nuclei counts. There is a distinct peak at around 15000-16000 number of parameters (nuclei).
	
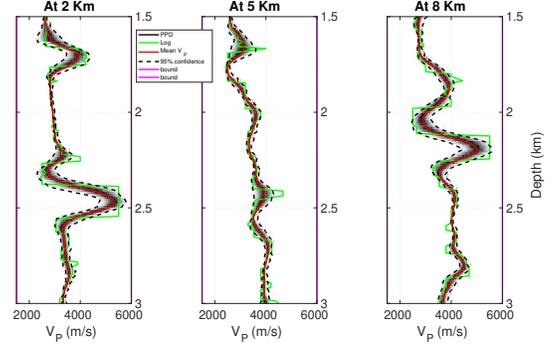
	\plot{locationPPD2_1}{width=\columnwidth}{Plot of marginal PPD (grey cloud), true log (green), mean parameter estimate (red), and 95\% confidence interval (dashed line) for well at location X=(2,5,8)Km marked with Green Location }
	
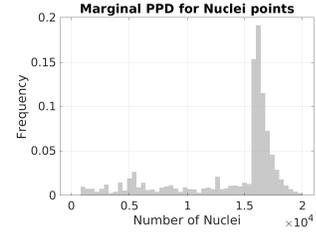
	\plot{kPPD15k1}{width=0.5\columnwidth}{Plot of marginal PPD for number of nuclei count}

	We observe that with depth, the uncertainty in the estimation of the P-wave velocity  increases. \Cref{fig:meanModel15k,fig:mapModel,fig:medianModel,fig:perstd15k1} show the mean model, maximum-a-posterioiri model, median model and the standard deviation (as a percentage of the mean model) at each grid. The maximum percentage standard deviation is nearly 10\%, and the standard deviation seems to be increasing with depth as the data coverage becomes poorer with depth. A Bayesian inversion is an essential tool for model exploration and uncertainty quantification. The sampled posterior distribution of the model parameters helps to investigate the quality and resolution power of the inversion. Comparing the prior distribution with the posterior distribution gives an idea about the recovered or improvement of resolution for the seismic model. In our experiment, we set the standard deviation in the Prior distribution large.  \Cref{fig:meanModel15k,fig:perstd15k1} display plots of posterior mean and posterior standard deviation specifically. Note that, we show posterior standard deviations relative to the corresponding mean values. Zones of low standard deviations correspond to regions of well-resolved velocities while velocities within zones of high standard deviation are poorly resolved. These are generally dictated by experimental geometry and ray-path coverage in the subsurface. For example, along the edges, the model resolution is poor, and the standard deviation is also significant in the deeper parts of the models. Note also that the standard deviation at the interface has a high relative standard deviation in the region. It is due to the  velocity-depth trade-off. It represents that position of the nuclei, and the parameter value is not entirely an independent quantity. To further investigate this observation, we plot model parameter correlation matrix for a small vertical and horizontal line containing around 45-50 grid points in \Cref{fig:verticalModelCovar,fig:horizontalModelCovar} respectively. It is clearly evident that in the vertical correlation matrix, as given in \Cref{fig:verticalModelCovar}, the model parameters are spread throughout and there is a very high correlation of a parameter location with its neighbors. The same positive correlation is not that prominent in the horizontal direction correlation matrix given in \Cref{fig:horizontalModelCovar}.
	
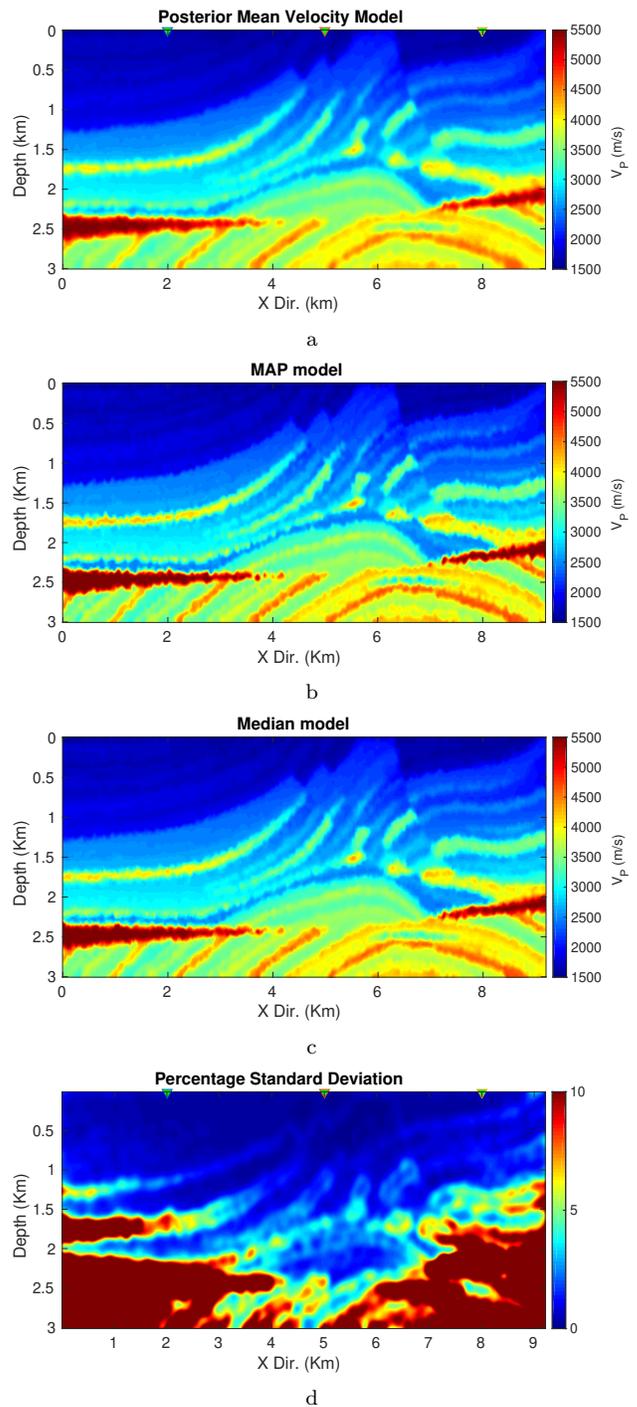
	\multiplot{4}{meanModel15k,mapModel,medianModel,perstd15k1}{width=\columnwidth}{Plot of the estimated (a) mean velocity model, (b) Maximum-a-posterior model, and (c) median model. (d) Plot of the percentage standard deviation of of the estimated velocity model.}

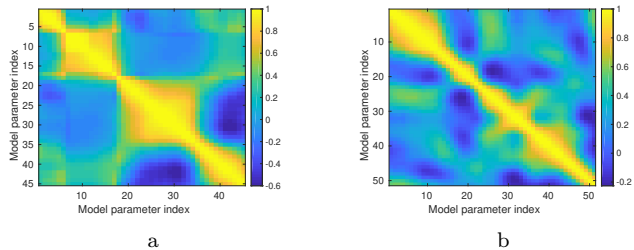
	\multiplot{2}{verticalModelCovar,horizontalModelCovar}{width=0.45\columnwidth}{(a) Plot of the neighbouring model parameter correlation matrix in (a) vertical direction (b) horizontal direction.}

	Finally, \Cref{fig:voronoiComparison} shows the comparison of the Voronoi cell distribution and density at the beginning of the inversion and the end of the inversion. We started with 1000 nuclei  and later at the end of the RJHMC we have around 15500 nuclei. 
	
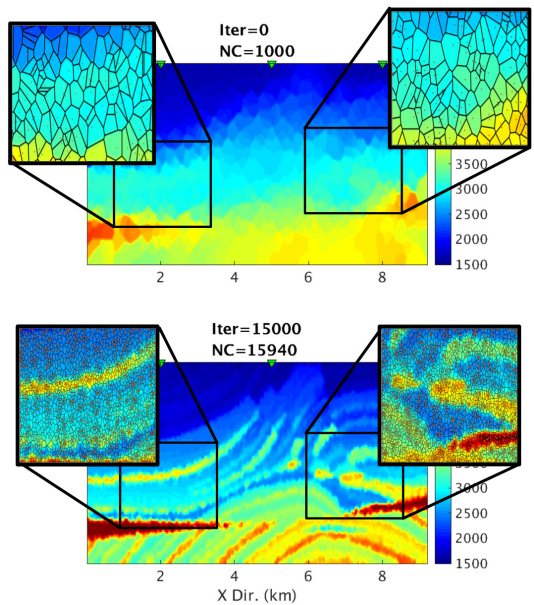
	\plot{voronoiComparison}{width=0.8\columnwidth}{Plot showing the comparison of the Voronoi cell distribution and density at the beginning of the inversion and at the end of the inversion. We started with 1000 nuclei count and later at the end of the RJHMC we have around 15500 nuclei count.}
	
	\section{Discussions}
	
	In this paper, we have demonstrated the applicability of a trans-dimensional MCMC method for a realistic 2D full waveform inversion problem. The main drawback of this method is the high computation and memory cost due to the cross-correlation of the forward and the adjoint state wavefields at the same time step. This requires saving the forward wavefields in the memory. Again, due to the limited GPU memory, data transfers are much slower compared to the computation, and the GPU waits for the data-copy to finish before continuing the computation. Therefore, we have followed a two-level checkpointing technique as given in \citet{datta2018asynchronous}, where the high overhead of memory usage is traded with the data movement and computations. For computation, we used four K40 Nvidia GPUs and NVMe (Non-Volatile Memory Express memory) at Institute for Geophysics, University of Texas at Austin. This enables us to have the data movement in the background, while the GPUs are continuously in use.

	As mentioned earlier, the Bayesian inversion has an intrinsic parsimony character~\citep{malinverno2002parsimonious}. Given the set of earth models having a different number of model parameters and fitting the data, models with fewer parameters are favored and have a higher posterior distribution. This feature can also be seen in our RJHMC result. In the first case, when we set the likelihood to unity, the sampled model is not influenced by the data. Thus, we can see in \Cref{fig:kPrior} all sampled models have equal probability. However, when the data influences the likelihood function, parametrization is dictated by a lower dimensional model due to the parsimonious nature of Bayesian inversion. This is evident in \Cref{fig:kPPD15k1}, where there is a peak at around 16000 nuclei, though a much higher model dimension is possible. This is around $8.5\%$ of the total nuclei possible (maintaining minimum separable distance). 
	
	Theoretically, the choice of parametrization used while solving an inverse problem does not prevent the model exploration to the target distribution if more than enough samples are drawn. However, for practical purposes, this sometimes cannot be achieved, and therefore different choice of parametrization produces different posterior distribution. In our application, Voronoi cells have been used for representing the velocity structure. However, there can be several other choices of parametrization which can be exploited depending on the problem. \citet{hawkinstrans} show a comparison of the result produced by three different types of parametrization: Voronoi, Delaunay triangulation with linear interpolants and with Clough-Tocher interpolant. They showed that for models with a sharp transition or discontinuities, Voronoi performs better than the other two. However, in case of a smoothly varying model, Delaunay triangulation with some interpolant would do better. Several sharp changes in velocity are expected in models like the Marmousi. Therefore, we choose Voronoi to represent our model over the other possibilities. Note that, since Voronoi cells can be discontinuous at the edges, which could produce spurious reflections in FWI. We apply smoothing to the velocity model before waveform simulations.

	\section{Conclusions}
	
	In this paper, we demonstrated the RJHMC algorithm and its application to a 2D FWI problem using the Marmousi model. We represented our 2D model using certain nuclei points. RJHMC algorithm solves the 2D FWI problem in a trans-dimensional way. It solves for the number of nuclei as well as the locations and velocities of the nuclei. The data dictate entirely the number of nuclei required, and is mainly determined automatically. The maximum number of nuclei possible is the dimension of the velocity grid, i.e., $251\times767=192517$, but in the example, using only $\approx15000-16000$ nuclei points, RJHMC is able to reconstruct the model. Using the gradient information from HMC, the algorithm can make some large jumps in the model space and efficiently sample the model space. RJHMC successfully reconstructed the complete model with an optimal number of model parameters as needed by the data. Multiple models sampled from the PPD can quantify uncertainty in the estimation of the P-wave velocity using 2D FWI.

	\bibliographystyle{seg}  
	\bibliography{bibliography}
\end{document}